\documentclass[10pt]{article}

\usepackage{epsf}
\usepackage{amsmath}
\usepackage{amsfonts}
\usepackage{amssymb} 
\usepackage{bm}
\usepackage{bbm}
\usepackage{graphicx}
\usepackage{epsfig}
\usepackage{latexsym}
\usepackage{nicefrac}
\usepackage{color}
\usepackage{cmbright}
\usepackage[latin2]{inputenc}

\DeclareSymbolFont{operators}   {OT1}{cmr} {m}{n}
\DeclareSymbolFont{letters}     {OML}{cmm} {m}{it}
\DeclareSymbolFont{symbols}     {OMS}{cmsy}{m}{n}
\DeclareSymbolFont{largesymbols}{OMX}{cmex}{m}{n}
\SetSymbolFont{operators}{bold}{OT1}{cmr} {bx}{n}
\SetSymbolFont{letters}  {bold}{OML}{cmm} {b}{it}
\SetSymbolFont{symbols}  {bold}{OMS}{cmsy}{b}{n}
\DeclareSymbolFontAlphabet{\mathrm}    {operators}
\DeclareSymbolFontAlphabet{\mathnormal}{letters}
\DeclareSymbolFontAlphabet{\mathcal}   {symbols}
\DeclareMathAlphabet      {\mathbf}{OT1}{cmr}{bx}{n}
\DeclareMathAlphabet      {\mathsf}{OT1}{cmss}{m}{n}
\DeclareMathAlphabet      {\mathit}{OT1}{cmr}{m}{it}
\DeclareMathAlphabet      {\mathtt}{OT1}{cmtt}{m}{n}
\SetMathAlphabet\mathsf{bold}{OT1}{cmss}{bx}{n}
\SetMathAlphabet\mathit{bold}{OT1}{cmr}{bx}{it}

\newcommand{\dd}{\mathrm{d}}
\newcommand{\ii}{\mathrm{i}}

\def\GG{{G}}

\def\calO{{\mathcal{O}}}

\def\uu{{u}}
\def\UU{{\mathcal{U}}}
\def\VV{{\mathcal{V}}}
\def\WW{{\mathcal{W}}}
\def\CC{{\mathcal{C}}}
\def\TT{{\mathcal{T}}}
\def\SW{{\mathrm{S}}}
\def\SW{{\mathrm{S}}}

\newcommand{\rept}{{\mathrm{Re}}}
\newcommand{\impt}{{\mathrm{Im}}}

\def\tfrac#1#2{ {\textstyle{\frac{#1}{#2}} } }
\def\half{   {\textstyle{\frac12}}   }

\newcommand{\PT}{${\mathcal{PT}}$}

\begin{document}

\vspace*{0.0cm}
\begin{center}
\begin{tabular}{c}
\hline
\rule[-3mm]{0mm}{12mm}
{\large \sf Calculation of the Characteristic Functions of Anharmonic Oscillators}\\
\hline
\end{tabular}
\end{center}
\vspace{0.0cm}
\begin{center}
Ulrich D. Jentschura\\
\vspace{0.2cm}
\scriptsize
{\em
Department of Physics,
Missouri University of Science and Technology\\
Rolla, Missouri, MO65409, USA}
\end{center}

\begin{center}
Jean Zinn--Justin\\
\vspace{0.2cm}
\scriptsize
{\em CEA, IRFU and Institut de Physique Th\'{e}orique,}\\
{\em Centre de Saclay, F-91191 Gif-Sur-Yvette, France}
\end{center}
\vspace{0.3cm}
\begin{center}
\begin{minipage}{14.0cm}
{\underline{Abstract}}
The energy levels of quantum systems are determined by quantization conditions.
For one-dimensional anharmonic oscillators, one can transform the
Schr\"{o}dinger equation into a Riccati form, i.e., in terms of the logarithmic
derivative of the wave function. A perturbative expansion of the logarithmic
derivative of the wave function can easily be obtained. The Bohr--Sommerfeld
quantization condition can be expressed in terms of a contour integral around
the poles of the logarithmic derivative. Its functional form is $B_m(E,g) = n +
\tfrac12$, where $B$ is a characteristic function of the anharmonic oscillator
of degree $m$, $E$ is the resonance energy, and $g$ is the coupling constant.
A recursive scheme can be devised which facilitates the evaluation of
higher-order Wentzel--Kramers--Brioullin (WKB) approximants.  The WKB expansion
of the logarithmic derivative of the wave function has a cut in the tunneling
region. The contour integral about the tunneling region yields the instanton
action plus corrections, summarized in a second characteristic function
$A_m(E,g)$. The evaluation of $A_m(E,g)$ by the method of asymptotic matching
is discussed for the case of the cubic oscillator of degree $m=3$.
\end{minipage}
\end{center}

\vspace{0.6cm}

Semiclassical techniques including WKB and Maslov methods;\\
General quantum mechanics and problems of quantization\\


\typeout{==> }
\typeout{==> Section: Introduction}
\typeout{==> }
%
%
\section{Introduction}
\label{intro}

The cubic anharmonic oscillator, defined by the Hamiltonian,
\begin{equation}
\label{h3}
h_3(g) = -\frac{1}{2} \,
\frac{\partial^2}{\partial q^2}  + \frac{1}{2} \, q^2 + 
\sqrt{g} \, q^3 \,,
\end{equation}
is a paradigmatic example of a quantum mechanical problem
which gives rise to complex resonance energies.
A particle initially trapped in the region $q \approx 0$
may tunnel through the classically forbidden well and 
escape toward $q \to -\infty$. For positive coupling $g > 0$,
the energies can be determined numerically by the method of
complex scaling\cite{BaCo1971,YaEtAl1978}. One scales the coordinate as
$q \to q \, {\rm e}^{{\rm i}\, \theta}$
for the cubic oscillator, which results in the
Hamiltonian
\begin{eqnarray}
\label{h3scaled}
H_{\rm c}(\theta) &=& 
{\rm e}^{-2i\theta}
\left( - \frac{1}{2} \frac{\partial^2}{\partial q^2} + 
\frac12 \, q^2 \, {\rm e}^{4 \ii \theta} + 
\sqrt{g} \, q^3 {\rm e}^{5 \ii \theta} \right) \,.
\end{eqnarray}
If one diagonalizes this complex operator 
in the basis of harmonic oscillator wavefunctions
$\{ \phi_n(q) \}_{n = 0}^{N_{\rm max}}$,
for large enough $N_{\rm max}$, 
one can numerically  determine the (complex) resonance energies of the
original cubic Hamiltonian~\eqref{h3}.
As discussed in Refs.~\cite{Al1988,YaEtAl1978},
these resonance energies are independent of $\theta$,
provided we choose $\theta$ sufficiently large
so that the rotated branch passes the position of the resonance
under investigation.
The numerical variation of the resonance energies 
of the cubic oscillator,
\begin{equation}
\epsilon^{(3)}_n(g) = \rept \left( \epsilon^{(3)}_n(g) \right) + 
\ii \, \impt \left( \epsilon^{(3)}_n(g) \right)\,,
\end{equation}
as a function of $g$ under a suitable increase
of $N_{\rm max}$ can be used to estimate the numerical
uncertainty of the numerical results
for the complex resonance energies~\cite{JeSuLuZJ2008}. 
Throughout this article, 
we keep the superscript ``3'' of the resonance energy 
$\epsilon^{(3)}_n(g)$ of the cubic 
oscillator in order to remain consistent with the 
notation of a brief Letter (see Ref.~\cite{JeSuZJ2009prl}), 
in which part of the results presented here have already 
been indicated.

For reasons which will explained below, we 
associate the resonance energy (with a negative imaginary part) 
with a value of the coupling parameter $g$ that has an infinitesimal positive
imaginary part, whereas a coupling with an infinitesimal
negative imaginary part is associated with an antiresonance,
that has a positive imaginary part. 
The resonance energies of the cubic potential as a function
of $g$ have a branch cut along the positive real axis,
and fulfill the dispersion relation~\cite{BeDu1999,BeWe2001},
\begin{equation}
\label{dispersionOdd}
\epsilon_n^{(M)}(g) = n + \frac12 + 
\frac{g}{\pi}\, \int_0^{\infty} \dd s \,
\frac{\impt \, \epsilon_n^{(M)}(s + {\rm i}\,0)}{s\, (s - g)}\,.
\end{equation}
We note that the spectrum of resonances is invariant under
the transformation $\sqrt{g} \to -\sqrt{g}$ in Eq.~\eqref{h3},
which can be compensated by a parity transformation $q \to -q$
(the latter leaves the spectrum manifestly invariant).

For weak positive coupling $g = |g| \to 0$ (with 
$\impt \, g > 0$), the imaginary part of the 
resonance energies is determined by a nonanalytic 
factor, which reads~\cite{JeSuLuZJ2008}
\begin{equation}
\label{imweak}
\impt \, \epsilon^{(3)}_{0}(g + \ii \,0) \sim
-\frac{2^{3n}}{n! \sqrt{\pi}} \, g^{-n-1/2} \, 
\exp\left(-\frac{2}{15\,g}\right)  
\qquad
\mbox{for}
\qquad
g = |g| \to 0 \,, \qquad
\impt \, g > 0 \,.
\end{equation}
For large positive coupling $g = |g| \to \infty + \ii \, 0 $, 
the first three resonance energies 
possess the following asymptotics [see Eq.~(17) and Table~II of 
Ref.~\cite{JeSuLuZJ2008}],
\begin{align}
\label{imstrong}
\epsilon^{(3)}_{0}(g + \ii \,0) 
\sim & \; 
g^{1/5} \, (0.617~160~050 - \ii \, 0.448~393~023) 
\qquad
\mbox{for}
\qquad
g = |g| \to \infty + \ii \, 0\,,
\nonumber\\[0.27ex]
\epsilon^{(3)}_{1}(g + \ii \,0) 
\sim & \; 
g^{1/5} \, (2.193~309~731 - \ii \, 1.593~532~797) 
\qquad
\mbox{for}
\qquad
g = |g| \to \infty + \ii \, 0\,,
\qquad
\nonumber\\[0.27ex]
\epsilon^{(3)}_{2}(g + \ii \,0) 
\sim & \; 
g^{1/5} \, (4.036~380~020 - \ii \, 2.932~601~744) 
\qquad
\mbox{for}
\qquad
g = |g| \to \infty + \ii \, 0\,.
\end{align}
All of the above results have a complex argument 
of $-\pi/5$, as they should. For intermediate $g$, a large number of 
resonance energies can be determined numerically by a suitable 
basis-set method and be used in order to construct
a complex-scaled time propagation algorithm for a quantum
mechanical wave packet (see Sec.~III of Ref.~\cite{JeSuLuZJ2008}).

A second available method for the calculation of 
resonance energies consists in the resummation of the divergent 
perturbation theory in complex directions of the parameters.
The perturbative coefficients
$\epsilon^{(3)}_{n,K}$ in the perturbation series
\begin{equation}
\label{weak}
\epsilon^{(3)}_{n}(g)
\sim
\sum_{K=0}^\infty \epsilon^{(3)}_{n,K} \, g^K \,, \qquad 
g \to 0\,,
\end{equation}
which describe the $n$th resonance energy of the cubic oscillator, 
diverge factorially for large perturbation theory order $K$.
Indeed, for large $K$, the leading asymptotics read 
(see, e.g., Ref.~\cite{JeSuZJ2009prl}),
\begin{equation}
\label{asymp}
\epsilon^{(3)}_{n,K}
\sim
- \frac{2^{2 n - 1/2 - K} \, 15^{n + K + 1/2}}{\pi^{3/2} \, n!} \,
\Gamma\left(n + K + \tfrac12\right) \,, \qquad
K \to \infty\,.
\end{equation}
With the exception of the coefficient of order zero, which 
reads $\epsilon^{(3)}_{n,0} = n + 1/2$, all $\epsilon^{(3)}_{n,K}$
are negative for $K \geq 1$. For $g > 0$, 
the series~\eqref{weak} is formally not Borel summable.
However, numerical evidence~\cite{CaEtAl2007,FrGrSi1985,Je2000prd,JeSuLuZJ2008}
suggests that the Borel--Pad\'{e} summation method with the 
Laplace--Borel integration being carried out along complex contours, as given by
Eqs.~(196)--(198) of Ref.~\cite{CaEtAl2007}, 
provides useful numerical approximants to the resonance energies.
The method has 
been put on rigorous mathematical grounds recently~\cite{Ca2000}
in the framework of distributional Borel summability~\cite{CaGrMa1986,CaGrMa1993}.
When the  integration contour $C_{+1}$ in the conventions
of Ref.~\cite{CaEtAl2007} is employed,
a negative sign is obtained for the imaginary part of the resonance energy.
The accuracy obtainable using 
complex Borel summation of the weak-coupling 
expansion~\eqref{weak} is restricted
by numerical oscillations in the values of the complex 
Borel--Pad\'{e} transforms as the order of the Pad\'{e} transformation is 
increased. These oscillations cannot be overcome even if extended-precision arithmetic
is used in intermediate steps of the calculation.
These oscillations have been described in
Refs.~\cite{JeSuLuZJ2008,SuLuZJJe2006} for the 
complex Borel--Pad\'{e} resummation method. 
Indeed, it has been observed in~\cite{JeSuLuZJ2008} that,
at a relatively moderate coupling $g = 0.6$,
the ground-state energy of the cubic oscillator as
determined by resummation cannot be calculated to better
accuracy than
\begin{equation}
\label{osc}
\epsilon^{(3)}_0(g = 0.6) = 0.554(1)- 0.351(6) \,{\rm i}
\end{equation}
by Borel resummation. The oscillations, which are described by the 
numerical uncertainty in the last digit reported in result
of Eq.~\eqref{osc}, cannot be overcome when the (Borel)
transformation order is increased and appear to represent
a fundamental limit of the convergence of resummed weak-coupling
perturbation theory in the case of a moderate and large (modulus of the) coupling
parameter $g$. A more accurate result obtained by complex scaling~\eqref{h3scaled} 
for the same coupling parameter is
\begin{equation}
\epsilon^{(3)}_0(g = 0.6) = 
0.554\,053\,519 - 0.351\,401\,778 \,{\rm i}\,.
\end{equation}
The applicability of the weak-coupling expansion for the 
cubic oscillator resonance energies thus is found to be severely limited
in the domain of moderate and large coupling, even if the 
perturbative expansion is {\em a posteriori} enhanced by a 
summation method acting in the complex plane.
Likewise, the leading-order analytic result~\eqref{imweak}
cannot be applied to the domain of moderate and large coupling $g$.
In this article, we explore the question whether one can find higher-order 
analytic formulas for the 
corrections to the leading result~\eqref{imweak} for the decay width of 
the ground-state resonance energy (and, possibly,
an arbitrary excited resonance energy) of the cubic oscillator,
and if yes, how large the correction terms are.

Indeed, in Ref.~\cite{JeSuZJ2009prl}, an analysis has been presented which 
allows us to calculate higher-order correction terms to the 
resonance energies for even and odd anharmonic oscillators
of arbitrary degree, and to formulate modified Bohr--Sommerfeld
quantization conditions for anharmonic oscillators such as the 
cubic one. Here, we provide a few more details of the analysis
that leads to the results of Ref.~\cite{JeSuZJ2009prl}, with 
a special emphasis on the cubic potential. To this end, we review
in Sec.~\ref{perturbative} the calculation of higher-order perturbative
(in $g$) approximants to the wave function
for the cubic potential. In Sec.~\ref{WKB}, the calculation of
higher-order corrections to the WKB approximation of the wave function
is reviewed. The modified quantization condition pertaining to 
the cubic oscillator is described in Sec.~\ref{modquant},
with a few new results for higher-order terms reserved for Sec.~\ref{results}.
Finally, conclusions are drawn in Sec.~\ref{conclu}.

\typeout{==> }
\typeout{==> Section: Perturbative Expansion}
\typeout{==> }
%
%
\section{Perturbative expansion}
\label{perturbative}

The Schr\"{o}dinger equation corresponding to the 
Hamiltonian given in Eq.~\eqref{h3} reads
\begin{equation}
\left( -\frac12 \, \frac{\partial^2}{\partial q^2} + 
\frac12 \, q^2 + \sqrt{g}\, q^3 \right) \, 
\varphi = E \, \varphi \,.
\end{equation}
If we transform the coordinate
according to $q \to q/\sqrt{g}$, we obtain
\begin{equation}
\label{hMscaled}
\left( -\frac{g^2}{2}\, \frac{\partial}{\partial q^2} + 
\VV(q) \right) \, \varphi = g \, E \, \varphi \,,
\qquad
\VV(q) = \frac12 \, q^2 + q^3 \,.
\end{equation}
We may transform to the Riccati equation by setting
\begin{equation}
\label{WKBprep}
\frac{\varphi'(q)}{\varphi(q)} = - \frac{s(q)}{g} \,,
\qquad
\frac{\varphi''(q)}{\varphi(q)} = 
\frac{s^2(q)}{g^2} - \frac{s'(q)}{g} \,,
\end{equation}
The Riccati form of the Schr\"{o}dinger equation then reads,
\begin{equation}
\label{riccati}
g \, s'(q) - s^2(q) + \UU^2(q) = 0 \,, \qquad
\UU(q) = \sqrt{ 2 \left[ \, \VV(q) - g \, E \, \right] }\,.
\end{equation}
The zeroth-order term is
\begin{equation}
s(q) \approx s_0(q) = \uu(q) = \sqrt{ 2 \, \VV(q) }  \,.
\end{equation}
The following recursion yields higher-order approximants
(in $g$) for the logarithmic derivative of the wave function,
\begin{align}
\label{system_s}
s(q) =& \; \sum_{K=0}^\infty g^K \, s_K(q) \,, \qquad
s_0(q) = \uu(q) \,, \qquad
s_1(q) = \frac{\uu'(q) - 2 E}{2 \uu(q)} \,,
\nonumber\\[1.377ex]
s_K(q) =& \; \frac{1}{2 \, \uu(q)} \left( s'_{K-1}(q) - 
\sum_{l = 1}^{K-1} s_{K-l}(q) \, s_l(q) \right) \,, 
\qquad \qquad K \geq 2.
\end{align}
We can divide $s(q)$ into a symmetric and an antisymmetric 
component under the simultaneous interchange
$g \to -g$, $E \to -E$,
\begin{equation}
s(q) = s_+(q) + s_-(q) = s_+(q, E, g) + s_-(q, E, g) \,,  \qquad
s_\pm(q, -E, -g) = \pm s_\pm(q, E, g) \,.
\end{equation}
From the following system of equations,
\begin{subequations}
\label{system}
\begin{align}
\label{systema}
g \, s'_-(q) - s_+(q)^2 - s_-(q)^2 +  \UU^2(q) =&\; 0 \,,
\\[1.377ex]
\label{systemb}
g \, s'_+(q) - 2 \, s_+(q) \, s_-(q) =&\; 0 \,,
\end{align}
\end{subequations}
we can eliminate the antisymmetric component by setting
\begin{equation}
s_-(q) = \frac{g}{2} \, \frac{s'_+(q)}{s_+(q)} \,.
\end{equation}
The Bohr--Sommerfeld quantization condition reads
\begin{equation}
\label{integral1}
- \frac{1}{2 \pi \ii \, g} \, \oint_C \dd z\, s_+(z) = n + \frac12 \,,
\end{equation}
where $C$ is a contour that encloses the zeros of the function 
$s_+(z)$ in the anticlockwise (mathematically positive) 
direction, and $n$ is the quantum number of the level. 
The function $B_3(E, g)$, defined by 
\begin{equation}
\label{BMEG}
B_3(E,g) \equiv - \frac{1}{2 \pi \ii \, g} \, 
\oint_C \dd z\, s_+(z) \,,
\end{equation}
[we recall the implicit dependence of $s_+(z)$ on $g$ and $E$]
provides for a universal means of determining the perturbative expansion
for an arbitrary excited level of the cubic potential, by means of the perturbative 
quantization condition $B_3(E,\GG) = n + \frac12$.
A calculation leads to the result
\begin{equation}
\label{B3}
B_3(E, g) = E + g \left( \frac{7}{16} + 
\frac{15}{4} E^2 \right) 
+ g^2 \left( \frac{1365}{64} \, E + \frac{1155}{16} \, E^3 \right) 
+ \, g^3 \left( \frac{119119}{2048} + 
\frac{285285}{256} E^2 + \frac{255255}{128} E^4 \right) + \mathcal{O}(g^4) .
\end{equation}

\typeout{==> }
\typeout{==> Section: WKB Expansion and Contour Integral}
\typeout{==> }
%
%
\section{WKB expansion and contour integral}
\label{WKB}

The WKB expansion is an expansion in $g$ at $g\,E$ fixed, 
i.e., it implies large values of $E$. We use this 
expansion in order to derive, roughly speaking, higher-order 
corrections to the instanton action which can later be used in 
order to derive higher-order corrections to resonance energies $E$
of levels whose principal quantum number is {\em not} large 
(this is based on a rather fortunate enhancement of the applicability 
regions of certain expansions, as explained below). 
The instanton configuration for the cubic potential has been 
discussed in Ref.~\cite{JeSuZJ2009sigma}, with a 
graphical representation being given in Fig.~3(b) of~\cite{JeSuZJ2009sigma}.
In order to achieve an instanton configuration pertaining to
positive values of the coordinates, we scale 
the coordinate as $q \to -q/\sqrt{g}$. We start from
\begin{equation}
\left( -\frac12 \, \frac{\partial^2}{\partial q^2} + 
\frac12 \, q^2 + g \, q^3 \right) \, 
\phi = E \, \phi
\end{equation}
and transform according to $q \to -q/\sqrt{g}$,
which gives
\begin{equation}
\label{hMscaled2}
\left( -\frac{g^2}{2}\, \frac{\partial}{\partial q^2} + 
\WW(q) \right) \, \phi = g \, E \, \phi \,, 
\qquad
\WW(q) = \frac12 \, q^2 - q^3 \,.
\end{equation}
We note that $\WW(q)$ differs from $\VV(q)$ in the sign of the 
cubic term. Setting
\begin{equation}
\frac{\phi'}{\phi} = - \frac{\SW(q)}{g} \,,
\end{equation}
where the logarithmic derivative is denoted 
$\SW(q)$ instead of $s(q)$, we obtain
\begin{equation}
\label{riccatiW}
g \, \SW'(q) - \SW^2(q) + \TT^2(q) = 0 \,, \qquad
\TT(q) = \sqrt{ 2 \left[ \, \WW(q) - g \, E \, \right] }\,.
\end{equation}
The zeroth-order term is different as compared to the 
perturbative expansion,
\begin{equation}
\SW(q) \approx \SW_0(q) = 
\sqrt{ 2 \left[ \, \WW(q) - g \, E \, \right] } \,.
\end{equation}
A recursion in $g$ can be defined,
\begin{align}
\label{SWrecur}
\SW(q) =& \;  \sum_{K=0}^\infty g^K \, \SW_K(q) \,, 
\qquad \SW_0(q) = \TT(q) \,, \qquad
\SW_1(q) = \frac{\SW_0'(q)}{2 \, \SW_0(q)} \,,
\nonumber\\[1.377ex]
\SW_K(q) =& \; \frac{1}{2 \, \SW_0(q)} \left( \SW'_{K-1}(q) - 
\sum_{l = 1}^{K-1} \SW_{K-l}(q) \, \SW_l(q) \right) \,.
\end{align}
Like the perturbative expansion, the WKB expansion of the logarithmic 
derivative of the wave function can be calculated recursively in 
ascending powers of $g$, and we can also distinguish between 
even and odd components under the symmetry operator 
$g \to -g$, $E \to -E$,
\begin{equation}
\label{evenodd2}
\SW(q) = \SW_+(q) + \SW_-(q) = \SW_+(q, E, g) + \SW_-(q, E, g) \,, 
\qquad
\SW_\pm(q, -E, -g) = \pm \, \SW_\pm(q, E, g) \,.
\end{equation}
The contour 
integral about the cut of the function $\SW_+(q)$  leads to 
a more complex structure as compared to 
the right-hand side of Eq.~(\ref{integral1}).
It reads
\begin{equation}
\label{WKBintegral}
\frac{1}{g} \; \oint\limits_{\CC} \dd z \; \SW_+(q) = 
A_3(E, g) + \tfrac12 \ln(2 \pi) 
- \ln \left[ \Gamma\left( \half - B_3(E, g) \right) \right]
+ B_3(E, g) \, \ln\left( - \frac{g}{8} \right) \,.
\end{equation}
Here, $\CC$ is a contour that encloses the cut of the WKB
expansion of the logarithmic derivative of the wave function 
in the clockwise sense.
We thus have to calculate the integral
\begin{equation}
\label{defT}
T = \frac{1}{g} \oint\limits_{\cal C} \dd q\, \SW_+(q) \approx 
\frac{1}{g} \oint\limits_{\cal C} \dd q\, 
\left[ 
\SW_0(q) + g^2 \, \SW_2(q) + g^4 \, \SW_4(q) 
\right] =
T_0 + T_2 + T_4 \,,
\end{equation}
Here, the WKB approximants
$\SW_0(q)$, $\SW_2(q)$ and $\SW_4(q)$ are defined in Eq.~\eqref{SWrecur},
and the integrals $T_0$, $T_2$ and $T_4$ correspond to the 
contour integrals of the WKB approximants of the respective order. 
The zeroth-order term is
\begin{equation}
T_0 = \frac{1}{g} \oint\limits_{\cal C} \dd q\, \SW_0(q) 
\end{equation}
which involves the zeroth-order WKB approximant
\begin{equation}
\SW_0(q) = \sqrt{ 2 \WW(q) - g E } = \sqrt{ q^2 - 2 q^3 - g E } \,.
\end{equation}
If we define the square root function to have its
branch cut along the positive real axis, then the 
above expression has its branch cut (in the limit 
$g\,E \to 0$) from $q = 0$ to $q= \tfrac12$.
Directly above the cut, the value of the square root is positive,
while directly below, it is negative.
As the contour encircles the cut in the clockwise
direction, we have an integration interval
from zero to $\tfrac12$ above the real axis, while below the real axis,
we go from $\tfrac12$ to zero. With this definition, the
contour integral is
\begin{equation}
\label{defT0}
T_0 = \frac{1}{g} \oint\limits_{\cal C} \dd q\,
\sqrt{ q^2 - 2 q^3 - g E }
= \frac{2}{g} \, \rept
\int\limits_0^{1/2} \dd q\, ( q^2 - 2 q^3 - g E )^{1/2} \,.
\end{equation}
For $g \, E > 0$, the cut of the WKB approximant does not 
extend along the full interval $(0, \tfrac12)$ any more.
However, by specifying the real part of the final result,
we pick up only those terms which are due to the cut.

The method of asymptotic matching (see Ch.~7.4 of Ref.~\cite{BeOr1978})
can be used in order to evaluate~\eqref{defT0}.
It involves an overlapping parameter $\epsilon$,
that is also used in Lamb shift calculations~\cite{Fe1949,JePa1996,Pa1993} in order
to separate the low-energy from the high-energy contribution
to the bound-electron self-energy.
The overlapping parameter fulfills

\begin{equation}
0 < g\,E \ll \epsilon \,.
\end{equation}
We then separate the integration interval into three parts,
the first of which extends from $(0, \epsilon)$, 
the second, from $(\epsilon, \tfrac12 - \epsilon)$, 
and the third complementing interval finally reads
$(\tfrac12 - \epsilon, \tfrac12)$.

The method of evaluation depends on the particular 
integration interval.
For the interval $(0, \epsilon)$, we have to be careful
to avoid divergences for very small $q=0$.
We have to keep the expression
$\sqrt{ q^2 - 2 g\,E}$ within $\SW_0(q)$ in unexpanded form. However,
we can expand the expression
$\sqrt{ q^2 \, (1 - 2 q) - 2 g\,E} = \sqrt{ q^2 \, - 2 g\,E - 2 q^3}$
in the $q^3$ term, then do the $q$ integration, 
subsequently expand the result in $g\,E$, and finally in $\epsilon$
up to order $\epsilon^0$, keeping all divergent 
terms~\cite{Fe1949,JePa1996,Pa1993}.
This leads to the result 
\begin{align}
I_1 =& \; \rept \int\limits_0^{\epsilon} \dd q\, ( q^2 - 2 q^3 - g E )^{1/2}
= (g E) \, \left[ -\frac12 - 
\frac12 \, \ln\left( \frac{2 \epsilon^2}{g E} \right) \right]
+ (g E)^2 \, 
\left[ \frac{47}{16} + \frac{1}{4 \epsilon^2} + \frac{3}{2 \epsilon} -
\frac{15}{8} \, \ln\left( \frac{2 \epsilon^2}{g E} \right) \right]
\nonumber\\[0.27ex]
& \; + (g E)^3 \, \left[ \frac{13327}{192} 
+ \frac{1}{8 \epsilon^4} + \frac{5}{6 \epsilon^3} 
+ \frac{35}{8 \epsilon^2} + \frac{105}{4 \epsilon} -
\frac{1155}{32} \, \ln\left( \frac{2 \epsilon^2}{g E} \right) \right]
\nonumber\\[0.27ex]
& \; + (g E)^4 \, 
\left[ \frac{6364777}{3072} 
+ \frac{5}{48 \epsilon^6} 
+ \frac{7}{8 \epsilon^5} 
+ \frac{315}{16 \epsilon^4} 
+ \frac{385}{16 \epsilon^3} 
+ \frac{15015}{128 \epsilon^2} 
+ \frac{45045}{64 \epsilon} 
- \frac{255255}{256} \, \ln\left( \frac{2 \epsilon^2}{g E} \right) 
\right] + \calO(g\,E)^5 \,.
\end{align}
For the interval $(\epsilon, \tfrac12 - \epsilon)$, we can
expand in $g\,E$ to any power, because all divergences near 
$q=0$ or $q= 1/2$ are cut off by the overlapping $\epsilon$ parameter.
The result reads
\begin{align}
I_2 =& \; \rept \int\limits_{\epsilon}^{1/2 -\epsilon} 
\dd q\, ( q^2 - 2 q^3 - g E )^{1/2}
= \frac{1}{15} +
(g E) \, \left[ \frac12 \, \ln\left( \frac{\epsilon^2}{4} \right) \right]
+ (g E)^2 \,
\left[ \frac{47}{8}
- \frac{1}{4 \epsilon^2} 
- \frac{3}{2 \epsilon} 
- \frac{2 \sqrt{2}}{\sqrt{\epsilon}}
+ \frac{15}{8} \, \ln\left( \frac{\epsilon^2}{4} \right) \right]
\nonumber\\[0.27ex]
& \; + (g E)^3 \, \left[ \frac{23189}{192} 
- \frac{1}{8 \epsilon^4} - \frac{5}{6 \epsilon^3} 
- \frac{35}{8 \epsilon^2} - \frac{105}{4 \epsilon} 
- \frac{4 \sqrt{2}}{3 \epsilon^{3/2}}
- \frac{40 \sqrt{2}}{\epsilon^{1/2}}
+ \frac{1155}{32} \, \ln\left( \frac{\epsilon^2}{4} \right) \right]
+ (g E)^4 \, \left[ \frac{5241655}{1536} - \frac{5}{48 \epsilon^6} 
\right.
\nonumber\\[0.27ex]
& \; \left. \quad - \frac{7}{8 \epsilon^5} 
- \frac{315}{64 \epsilon^4} 
+ \frac{385}{16 \epsilon^3} 
- \frac{15015}{128 \epsilon^2}
- \frac{45045}{64 \epsilon}
- \frac{2 \sqrt{2}}{\epsilon^{5/2}}
- \frac{140 \sqrt{2}}{\epsilon^{3/2}}
- \frac{1120 \sqrt{2}}{\epsilon^{1/2}}
+ \frac{255255}{256} \, \ln\left( \frac{\epsilon^2}{4} \right) 
\right] + \calO(g\,E)^5 \,.
\end{align}
The result from the interval $I_3$ 
compensates the divergences of fractional order in $\epsilon$
from $I_2$,
\begin{align}
I_3 =& \; \rept \int\limits_{1/2-\epsilon}^{1/2} 
\dd q\, ( q^2 - 2 q^3 - g E )^{1/2}
= (g E)^2 \, \left[ \frac{2 \sqrt{2}}{\epsilon^{1/2}} \right]
+ (g E)^3 \, \left[ 
\frac{4 \sqrt{2}}{3 \epsilon^{3/2}} +
\frac{40 \sqrt{2}}{\epsilon^{1/2}} 
\right]
\nonumber\\[0.27ex]
& \; + (g E)^4 \, 
\left[ 
\frac{2 \sqrt{2}}{\epsilon^{5/2}}
+ \frac{140 \sqrt{2}}{3 \epsilon^{3/2}}
+ \frac{1120 \sqrt{2}}{\epsilon^{1/2}}
\right] + \calO(g\,E)^5 \,.
\end{align}
In the result for $T_0$,
\begin{equation}
T_0 = \frac{2}{g} \, \left( I_1 + I_2 + I_3 \right) \,,
\end{equation}
the overlapping parameter $\epsilon$ cancels, and we obtain
\begin{align}
T_0 =& \; \frac{2}{15 \, g} + \left\{ -E + E\,\ln\left( \frac{g E}{8} \right) \right\}
+ g \, \left[ - \frac{141}{8} E^2- 
\frac{15}{4} \, E^2 \, \ln\left( \frac{g E}{8} \right) \right]
\\[0.27ex]
& 
+ g^2 \, \left[ \frac{3043}{8} E^3 + 
\frac{1155}{16} \, E^3 \, \ln\left( \frac{g E}{8} \right) \right]
+ g^3 \, \left[ \frac{5616029}{512} E^4 +
\frac{255255}{128} \, E^4 \, \ln\left( \frac{g E}{8} \right) \right] \,.
\nonumber
\end{align}
The second term reads
\begin{align}
T_2 =& \; g \oint\limits_{\cal C} \dd q\, \SW_2(q) =
-\frac{1}{24\,E} + g \, \left[ \frac{41}{16} +  
\frac{7}{16} \, \ln\left( \frac{g E}{8} \right) \right]
\\[0.27ex]
& \; + g^2 \, \left[ \frac{16431}{128} E + 
\frac{1365}{64} \, E \, \ln\left( \frac{g E}{8} \right) \right]
+ g^3 \, \left[ \frac{862501}{128} E^2 -
\frac{285285}{256} \, E^2 \, \ln\left( \frac{g E}{8} \right) \right] \,.
\nonumber
\end{align}
The third term is 
\begin{equation}
T_4 = g^3 \oint\limits_{\cal C} \dd q\, \SW_4(q) =
\frac{7}{2880\,E^3} 
- \frac{7\,g}{768\,E^2} 
+ \frac{539\,g^2}{1024\,E} 
+ g^3 \, \left[ \frac{27101039}{611440} +
\frac{119119}{2048} \ln\left( \frac{g E}{8} \right) \right] \,.
\end{equation}
The total result of the contour integral of the WKB expansion is 
$T = T_0 + T_2 + T_4$, where
\begin{align}
& T = \frac{7}{2880\,E^3} - \frac{1}{24\,E} +
\frac{2}{15 g} + 
\left\{ - E + E \, \ln\left( \frac{g E}{8} \right) \right\}
+ g \, \left[ - \frac{7}{768\,E^2} + \frac{41}{16} + \frac{141}{8} E^2 
+ \left( \frac{7}{16} + \frac{15}{4} \, E^2 \right) \, 
\ln\left( \frac{g E}{8} \right) \right]
\nonumber\\[0.27ex]
& 
+ g^2 \, \left[  \frac{539}{1024\,E} + \frac{16431}{128} E + \frac{3043}{8} \, E^3 
+ \left( \frac{1365}{64} E + 
\frac{1155}{16} \, E^3 \right) \, \ln\left( \frac{g E}{8} \right) \right]
\nonumber\\[0.27ex]
& + g^3 \, \left[ \frac{27101039}{61440} + \frac{862501}{128} E^2 
+ \frac{5616029}{512} E^4 
+ \left( \frac{119119}{2048} + \frac{285285}{256} \, E^2 + 
\frac{255255}{128} E^4 \right) \, 
\ln\left( \frac{g E}{8} \right) \right] +
\calO(g^4) \,.
\end{align}
The perturbative counterterms read
\begin{align}
& P = \tfrac12 \ln(2 \pi) - \ln \left[ \Gamma\left( \half - B_3(E, g) \right) \right]
+ B_3(E, g) \, \ln\left( -\frac{g}{8} \right) 
= \frac{7}{2880\,E^3} - \frac{1}{24\,E} 
+ \left\{ - E + E \, \ln\left( \frac{g E}{8} \right) \right\}
\nonumber\\[0.27ex]
& + g \, 
\left[ - \frac{7\,g}{768\,E^2} + \frac{5}{32} 
+ \left( - \frac{7}{16} + 
\frac{15}{4} \, E^2 \right) \, \ln\left( \frac{g E}{8} \right) \right]
\nonumber\\[0.27ex]
& + g^2 \, \left[  \frac{539}{1024\,E} + \frac{65}{16} E + \frac{225}{32} \, E^3 
+ \left( \frac{1365}{64} E + 
\frac{1155}{16} \, E^3 \right) \, \ln\left( \frac{g E}{8} \right) \right]
\nonumber\\[0.27ex]
& + g^3 \, \left[ \frac{319081}{8192} + \frac{175325}{1024} E^2 
+ \frac{33525}{128} E^4 + \left(  \frac{119119}{2048} + \frac{285285}{256} \, E^2 + 
\frac{255255}{128} \, E^4 \right) \, \ln\left( \frac{g E}{8} \right) \right] \,.
\end{align}
where the result of $B_3(E,g)$ from Eq.~\eqref{B3}
has been used, as well as the asymptotic expansion
\begin{equation}
\label{lngamma}
\ln\Gamma( \tfrac12 + z) = 
z \, \left\{ \ln(z) -1 \right\} + \tfrac12 \ln(2\pi) - \frac{1}{24 \, z} + 
\frac{7}{2880\, z^3} + \dots \,.
\end{equation}
Note that the logarithm $\ln(g E/8)$ is obtained as the combination 
of $B_3(E, g) \, \ln(-g/8) \approx E\,\ln(-g/8) $ and from the asymptotic expansion
of the expression 
$-\ln\Gamma( \tfrac12 - B_3(E,g)) \approx E\, \ln(-E)$. 
Finally, the instanton function $A_3(E,g)$ for the cubic potential 
is found to read
\begin{align}
\label{A3}
A_3(E,g) =& \; T - P = \frac{2}{15\,g} + 
g\, \left( \frac{77}{32} + \frac{141}{8} \, E^2 \right) +
g^2\, \left( \frac{15911}{128} \, E + \frac{11947}{32} \, E^3 \right) 
\\[1.377ex]
& \; + g^3\, \left( \frac{49415863}{122880} +
\frac{6724683}{1024}\, E^2 +
\frac{5481929}{512}\, E^4 \right) + \calO(g^4) \,.
\end{align}
This result for the instanton function is the basis for the 
calculation of higher-order corrections to the decay width
of the resonance energies, as will be explained in the following.

%
%
\section{Modified quantization condition}
\label{modquant}

For $g < 0$, the interaction Hamiltonian 
$\sqrt{g} \, q^3 = \pm \ii \, | \sqrt{g} | \, q^3$  is
\PT{}-symmetric, and its spectrum is purely real.
In view of Eqs.~\eqref{integral1} and~\eqref{BMEG}, we 
then have the purely perturbative quantization condition
\begin{align}
B_3(E,g) =& \; n + \tfrac12 \quad \Leftrightarrow \quad
\frac{1}{\Gamma\left( \tfrac12 - B_3(E,g)\right)} = 0 \,, 
\qquad g < 0 \,.
\end{align}
Let us now study how this condition needs to be 
modified in case instantons exist, i.e., for $g > 0$.
We start from the unperturbed harmonic oscillator
\begin{equation}
{\cal H} = -\frac12 \, \frac{\partial^2}{\partial q^2} + \frac12 \, q^2 \,.
\end{equation}
In this form, we naturally identify
\begin{equation}
\frac{1}{\Gamma(\tfrac12 - E)} = {\rm det}({\cal H}-E) 
\end{equation}
as the spectral determinant. The effect of the instanton is to
add an infinitesimal imaginary part to the energy of the bound state,
\begin{equation}
E \approx {\cal E} \equiv n + \tfrac12 + \ii \, \impt \, {\cal E} \,.
\end{equation}
Indeed, the definition of resonance energies via complex scaling and the proof of
their existence can be found in Ref.~\cite{CaGrMa1980}.
Expanding in the nonperturbatively small (in $g$) imaginary part of the 
energy, we obtain
\begin{equation}
\label{seed}
\frac{1}{\Gamma(\tfrac12 - {\cal E})} \approx
- (-1)^{n+1/2} \; n! \;  \impt \, {\cal E} \,.
\end{equation}
For the cubic oscillator, we approximate the right-hand side of
Eq.~\eqref{seed} as
\begin{equation}
\label{gen_im_odd}
\impt \, {\cal E} \approx 
{\rm Im} \, \epsilon_n^{(M)}(g) \approx
-\frac{2^{3 n}}{n! \sqrt{\pi}} \, 
g^{-n-1/2} 
\, \exp\left( -\frac{2}{15 \, g} \right) 
\approx -\frac{1}{n! \sqrt{8 \pi}} \, 
\left( \frac{8}{g} \right)^{B_3(E,g)} 
\, \exp\left( -A_3(E,g) \right) \,.
\end{equation}
We now generalize the left-hand side of Eq.~\eqref{seed}
as $1/\Gamma(\tfrac12 - B_3(E,g))$ and approximate the
right-hand side of Eq.~\eqref{seed} by the right-hand side
of Eq.~\eqref{gen_im_odd}, and obtain
\begin{align}
\label{quantOdd}
& \frac{1}{\Gamma\left( \half - B_3(E, g)\right)} = 
\frac{1}{\sqrt{8 \pi}}\, 
\left( - \frac{8}{g} \right)^{B_3(E, g)} \,
\exp[-A_3(E, g)] \,, \qquad g > 0 \,,
\end{align}
which is our conjecture for the resonance energies 
of the cubic anharmonic oscillator in the 
``unstable'' regime of positive coupling parameter $g > 0$.
When the logarithm of both sides of Eq.~\eqref{quantOdd} is 
taken, we recover the structure of the terms on the 
right-hand side of Eq.~\eqref{WKBintegral}.

In Sec.~\ref{WKB}, we have calculated the 
function $A_3(E,g)$ in the WKB limit $g \to 0$,
with $g\,E$ fixed. 
In order to derive the result~\eqref{A3} for the 
function $A_3(E,g)$, we had to expand the 
function $B_3(E,g)$ in the limit of large $E$,
which implies [according to Eq.~\eqref{lngamma}]
the approximation $- \ln \Gamma( \tfrac12 - B_3(E,g) ) \approx
E \ln \Gamma( - E )$. {\em A priori}, this expansion is not applicable when 
$E \approx n + \tfrac12 + \delta$, where $\delta$ 
summarizes the perturbative and nonperturbative corrections
(in $g$). We have calculated the instanton function $A_3(E,g)$
in an ``unphysical'' domain of large quantum numbers $n$ (large $E$),
``away'' from the first few resonance  energies.
However, the result~\eqref{A3} is written in terms 
of an expansion whose terms actually decrease for 
typical values of $E$ for the first few resonance energies,
where $E$ is close to a small half-integer.
We can thus make an---in some sense fortunate---observation:
While the expansion~\eqref{lngamma} for the logarithm of the $\Gamma$ function is 
not applicable to the physical domain of $E = n + \tfrac12 + \delta$,
the obtained expansion for the $A_3(E,g)$ function is applicable
in both the ``unphysical'' WKB domain of $g\,E$ fixed, and $g\to 0$,
thus $E \to \infty$, as well as the physical domain of 
$E \approx n + \tfrac12$, and $g \to 0$.
We can thus use the modified quantization condition~\eqref{quantOdd}
in the regime of low principal quantum numbers, which is of prime interest.

This unified result, which gives immediate and systematic access
to the higher-order corrections,  should be compared
to other approaches (Refs.~\cite{HaKl2003,HaKl1994,Vo1999a,Vo1999b},
which also strive to go beyond the simple leading-order results for the
decay rates of the resonances. One approach~\cite{HaKl2003,HaKl1994}
uses variationally improved perturbation theory which leads
to a numerical improvement over the simple summation of the
first few perturbative terms in the strong-coupling domain,
but does not lead to a systematic prescription for calculation of
higher-order coefficients in the ``resurgent'' expansion.
The other approach~\cite{Vo1999a,Vo1999b} is based on a set of
quantization conditions derived as variations of the Bohr--Sommerfeld
condition.

\typeout{==> }
\typeout{==> Section: Results}
\typeout{==> }
%
%
\section{Results}
\label{results}

Based on the results presented in Eqs.~\eqref{B3},~\eqref{A3}
and~\eqref{quantOdd}, it is easy to derive 
higher-order corrections to the imaginary part of 
the first few resonance energies of the cubic potential, 
by entering into the condition~\eqref{quantOdd} with an 
ansatz 
\begin{equation}
\label{coeff}
{\rm Im} \, \epsilon_n^{(3)}(g) = 
-\frac{2^{3n}}{n! \sqrt{\pi}} \, g^{-n-1/2} \, 
\exp\left(-\frac{2}{15\,g}\right) \,
\left( 1 + a_n \, g + b_n \, g^2 + \dots \right) \,,
\qquad \mbox{for} \qquad g > 0 \,,
\end{equation}
and equating coefficients. The higher-order coefficients,
for the lowest four states, read
\begin{subequations}
\label{reshigher}
\begin{align}
\label{im3level0}
{\rm Im} \, \epsilon_0^{(3)}(g) = & \;
-\frac{\exp\left( - \frac{2}{15\,g} \right)}%
{\sqrt{\pi \, g}} \, 
\left\{ 1 - \frac{169}{16} \, g - 
\frac{ 44507 }{ 512 } \, g^2 - 
\frac{ 86071851 }{ 40960 } \, g^3 - 
\frac{ 189244716209 }{ 2621440 } \, g^4 +
\calO(g^5) \right\} \,,
\\[1ex]
\label{im3level1}
{\rm Im} \, \epsilon_1^{(3)}(g) = & \;
- \frac{ 8 \, {\rm e}^{-2/(15\,g)} }{ \sqrt{\pi} \, g^{3/2}} \,
\left\{ 
1 - \frac{853}{16} \, g + 
\frac{33349}{512} \, g^2 - 
\frac{395368511}{40960} \, g^3 - 
\frac{ 1788829864593 }{ 2621440 } \, g^4 +
\calO(g^5) \right\} \,,
\\[1ex]
\label{im3level2}
{\rm Im} \, \epsilon_2^{(3)}(g) =  & \;
- \frac{ 32 \, {\rm e}^{-2/(15\,g)} }{ \sqrt{\pi} \, g^{5/2}} \,
\left\{ 
1 - \frac{2101}{16} \, g 
+ \frac{1823341}{512} \, g^2 - 
\frac{1085785671}{40960} \, g^3 - 
\frac{4272925639361}{2621440} \, g^4 +
\calO(g^5) \right\} \,,
\\[1ex]
\label{im3level3}
{\rm Im} \, \epsilon_3^{(3)}(g) = & \;
- \frac{ 256 \, {\rm e}^{-2/(15\,g)} }{ 3 \sqrt{\pi} \, g^{7/2}} \,
\left\{ 
1 - \frac{3913}{16} \, g 
+ \frac{8807869}{512} \, g^2 - 
\frac{1571666861}{40960} \, g^3 - 
\frac{3214761534593}{2621440} \, g^4 +
\calO(g^5) \right\}
\end{align}
\end{subequations}
For the coefficient of relative order $g$, a general
result (for any $n$) has been indicated in Eq.~(22) of Ref.~\cite{Al1988}.
For $n=2$, Eq.~(22) of Ref.~\cite{Al1988} gives 
a correction term 
of $-2041 g/16$ instead of $-2101 g/16$ for the state with $n=2$,
and a correction term 
of $-10543 g/48$ instead of $-3913 g/16$ for the state with $n=3$.
With the help of the dispersion relation~\eqref{dispersionOdd}, 
these coefficients can be related to the corrections to the 
leading factorial asymptotics~\eqref{asymp} of the perturbative coefficients
of relative order $1/K$. 
We have checked our calculation against numerical values of the 
perturbative coefficients of up to 60th order, which can easily be determined
on the basis of the relation $B_3(E,g) = n + 1/2$.

\typeout{==> }
\typeout{==> Section: Conclusions}
\typeout{==> }
%
%
\section{Conclusions}
\label{conclu}

The calculation of higher-order corrections to the 
decay widths of quantum states in unstable potentials
is a challenging problem. Analytic approximations are 
useful in that regard, if they contain a sufficient number of 
correction terms which also permit an estimate of the 
truncation error.
As shown in Secs.~\ref{modquant} and~\ref{results}, the 
calculation of higher-order corrections to the 
decay widths of resonances of the cubic anharmonic oscillator, 
is based on two characteristic functions, namely, the 
\begin{equation}
\mbox{perturbative function} \qquad B_3(E,g) = E + \calO(g)  \qquad
\mbox{defined in Eq.~\eqref{BMEG} and evaluated in Eq.~\eqref{B3}}
\end{equation}
and on the 
\begin{equation}
\mbox{instanton function} \qquad A_3(E,g) = \frac{2}{15\,g} + \calO(g)  \qquad
\mbox{defined in Eq.~\eqref{WKBintegral} and evaluated in Eq.~\eqref{A3}}.
\end{equation}
The first of these functions is given by an evaluation of a contour 
integral about the poles of the perturbative approximation 
to the logarithmic derivative of the wave function (see Sec.~\ref{perturbative}).
The second function, $A_3(E,g)$, is determined by a contour integral
about the cut of the WKB approximation of the wave function (see Sec.~\ref{WKB}).

The modified Bohr--Sommerfeld quantization condition~\eqref{modquant}
then gives us access to the higher-order corrections to the 
imaginary part of the resonance energies. These are given, for the 
first few resonances of the cubic potential, in Sec.~\ref{results}.
The full solution of the modified quantization condition
has the structure of a generalized nonanalytic expansion, or 
``resurgent'' expansion,
\begin{align}
\label{genOddHigher}
& \epsilon_n^{(3)}(g) = \;
\sum_{J=0}^\infty 
\left[ 
\frac{\ii}{n! \, \sqrt{8 \pi}} \,
\left( \frac{8}{g} \right)^{n + \half} \,
\exp\left(-\frac{2}{15\,g}\right)
\right]^J
\sum_{L=0}^{L_\mathrm{max}}
\, \ln^L\left( -\frac{8}{g}  \right)
\sum_{K=0}^\infty \Xi^{(3,n)}_{J,L,K} \, g^{K} \,, \qquad g > 0 \,,
\end{align}
where 
\begin{equation}
L_\mathrm{max} = \mathrm{max}(0, J-1)\,.
\end{equation}
The perturbative coefficients $\epsilon^{(3)}_{n,K}$ are equal to 
the $\Xi^{(3,n)}_{0,0,K}$ coefficients.
The coefficients entering Eq.~\eqref{reshigher} are the 
$\Xi^{(3,n)}_{1,0,K}$ coefficients.

The coefficients $\Xi^{(3,n)}_{1,0,1}$
multiply the correction
of relative order $g$ to the one-instanton ($J=1$) effect 
and are given in Eq.~\eqref{reshigher} for $n=0,1,2,3$.
They grow quite drastically in magnitude
with the principal quantum number. In the notation of Eq.~\eqref{coeff},
we have for the state with $n=3$ a coefficient of
$a_3 = -244.563$. This means that the correction term of order
$g$ halves the total result for the decay width of the third
excited state of the cubic potential already at a minuscule coupling
parameter of $g \approx 0.002$.
It is quite interesting to see that numerically 
approximations to such a fundamental physical quantity
as the resonance energies of the cubic anharmonic 
oscillator require a considerable effort in their evaluation.
Here, our aim was to carefully explain the calculational approach 
necessary in order to gain access to the higher-order 
corrections, which are numerically significant. 

\section*{Acknowledgments}

The authors gratefully acknowledge support from the 
National Science Foundation (Grant PHY--8555454) and from the 
Missouri Research Board.

\end{document}